\documentclass[published]{nst}

\usepackage{subfigure,dcolumn}
\usepackage{epstopdf}
\usepackage{mhchem}
\usepackage{upgreek}
\usepackage{color}

\begin{document}

\title{Characterization of silicon microstrip sensors for space astronomy}
\thanks{Supported by the National Natural Science Foundation of China (No. 11873020, 11973097,  and U1738210), the Strategic Pioneer Program on Space Science of the Chinese Academy of Sciences (No. XDA15010200), and the National Key R\&D Program of China (No. 2016YFA0400204).}

\author{Jia-Ju Wei}
\email[Corresponding author, ]{weijj@pmo.ac.cn}
\affiliation{Purple Mountain Observatory, Chinese Academy of Sciences, Nanjing 210023, China}
\affiliation{\mbox{Key Laboratory of Dark Matter and Space Astronomy, Chinese Academy of Sciences,} Nanjing 210023, China}

\author{Jian-Hua Guo}
\affiliation{Purple Mountain Observatory, Chinese Academy of Sciences, Nanjing 210023, China}
\affiliation{\mbox{Key Laboratory of Dark Matter and Space Astronomy, Chinese Academy of Sciences,} Nanjing 210023, China}

\author{Yi-Ming Hu}
\affiliation{Purple Mountain Observatory, Chinese Academy of Sciences, Nanjing 210023, China}
\affiliation{\mbox{Key Laboratory of Dark Matter and Space Astronomy, Chinese Academy of Sciences,} Nanjing 210023, China}

\begin{abstract}

Silicon microstrip detectors are widely used in experiments for space astronomy. Before the detector is assembled, extensive characterization of the silicon microstrip sensors is indispensable and challenging. This work electrically evaluates a series of sensor parameters, including the depletion voltage, bias resistance, metal strip resistance, total leakage current, strip leakage current, coupling capacitance, and interstrip capacitance. Two methods are used to accurately measure the strip leakage current, and the test results match each other well. In measuring the coupling capacitance, we extract the correct value based on a SPICE  model and two-port network analysis. In addition, the expression of the measured bias resistance is deduced based on the SPICE model.

\end{abstract}


\maketitle

\section{Introduction}

$\Gamma$-ray detection is one of the most important observation methods in space astronomy. Recent studies showed that the high-energy $\gamma$-ray's spectrum and distribution can be used to explore the physical characteristics and distribution of the dark matter. By contrast, X-ray detection is also important in space astronomy. For example, hard X-ray solar observation can probe nonthermal electrons accelerated in the solar atmosphere. To accurately detect the $\gamma$-ray and X-ray distributions, a tracker with a very high spatial resolution is needed. Although there are many candidates \cite{d1,d2,d3,d4}, the silicon microstrip detector~\cite{c1,c2,c3,c4} is one of the most suitable candidates because it has excellent spatial resolution (lower than 1.8$\mu$m was mentioned in reference \cite{b1}). In addition, silicon microstrip detectors can extend the detection area by forming a cascade structure (i.e.,~the so-called ``ladder''~\cite{b19}) without adding extra readout electronics. This feature is very important for high-energy space astronomy detection, which needs a large detecting area but low power consumption.

Silicon microstrip detectors are widely used in experiments for space astronomy, e.g., Payload for Antimatter Exploration and Light-nuclei Astrophysics (PAMELA)~\cite{b2,b3}, Light Imager for $\gamma$-ray Astrophysics (AGILE)~\cite{b4,b5}, $\gamma$-ray Large Area Space Telescope (Fermi-GLAST)~\cite{b6,b7}, Alpha Magnetic Spectrometer (AMS-02)~\cite{b8,b9}, Dark Matter Particle Explorer (DAMPE)~\cite{b10,b11}, and Focusing Optics X-ray Solar Imager (FOXSI)~\cite{a1,a2}. Bases on DAMPE's experience, our collaborative group proposes a new generation of space telescope, i.e.,~Very Large Area gamma-ray Space Telescope (VLAST). In the preliminary design, 18 tracking planes, each consisting of two layers of single-sided silicon microstrip sensors, will be configured to measure the two orthogonal views perpendicular to the pointing direction of the detector.

Extensive characterization of the silicon microstrip sensors before assembling the detector is indispensable. Owing to the cascade structure, one bad sensor will deactivate the detecting ``ladder'' consisting of several sensors and generate a large ``dead'' area. By contrast, the characterization process is challenging. Owing to the on-chip parasitics, the silicon microstrip sensor acts as a complicated SPICE RC  network~\cite{b12,b13}, so it is challenging to measure or extract the right parameters. In addition, the actual values of some parameters cause difficulties with the measurement, e.g., very low current, resistance, or capacitance. This work evaluates the main sensor parameters, including the depletion voltage, bias resistance, metal strip resistance, total leakage current, strip leakage current, coupling capacitance, and interstrip capacitance.

\section{Basic sensor structure}

\begin{figure}[!htb]
\includegraphics[width=0.4\textwidth]{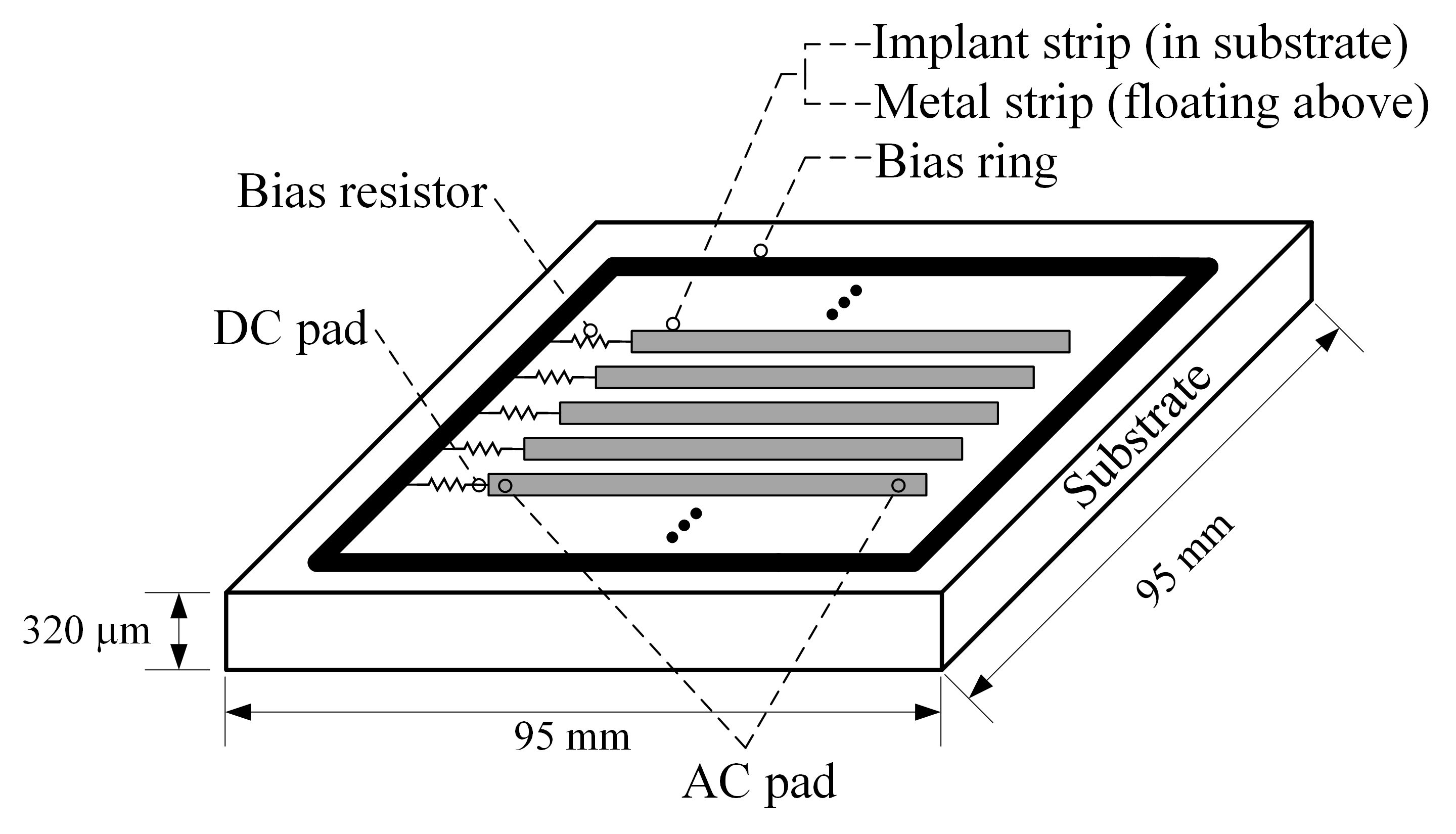}
\caption{Structure of silicon microstrip sensor}
\label{Fig_Structure-of-the-silicon-microstrip-sensor}
\end{figure}

In this work, we study a single-sided AC-coupled sensor (shown in Fig.~\ref{Fig_Structure-of-the-silicon-microstrip-sensor}) using a silicon wafer of 320 $\mu$m in thickness from Hamamatsu Photonics. The electrical resistivity of the N-doped silicon wafer is on the order of several k$\Omega\cdot$cm, and the chip size is 95 mm $\times$ 95 mm . The P$^+$ implant strips are 48 $\mu$m wide and 93 mm long, with a pitch of 121 $\mu$m. Each implant strip is biased with a poly-silicon resistor that is connected to a bias ring. The AC coupling is obtained using a silicon oxide layer between the implant strip and the floating metal strip. To prevent microdischarge at the edges, the metal strip covers the implant strip edges with a 4-$\mu$m overhang~\cite{b14}. In addition, the AC pads and DC pads of the sensor allow testing access to the metal strip and the implant strip, respectively.

\section{Characterization of sensor}

Because silicon microstrip sensors are very fragile and sensitive to contamination during transport, storage, and picking, we must evaluate them extensively before assembling them onto the hybrid board. By contrast, some sensor parameters may deviate from the nominal value by a range, so accurate measurements are needed to determine the actual value. In addition, the characterization process can improve our understanding of the sensor's working mechanism.

\subsection{Depletion voltage}

A silicon microstrip sensor consists of hundreds of strips, and each P$^+$ implant strip together with the N-doped substrate forms a typical approximate PN junction. When a reverse bias voltage is applied to the sensor, a depletion layer free of mobile carriers appears. The depletion voltage is the bias voltage, which extends the depletion layer to the thickness of the sensor.

According to Hamamatsu Photonics, the wafer resistivity of the silicon microstrip sensor is in the range of 5 to 10 k$\Omega\cdot$cm. We can roughly estimate the depletion voltage~\cite{b15} by
\begin{equation}
\label{Eq_Depletion-voltage-A}
V_{depl} = \frac{L^2}{2 \rho  \epsilon  \epsilon _0 \mu _e},
\end{equation}
where $L$ is the wafer thickness, $\mu_e$ is the electron mobility, $\rho$ is the wafer resistivity, $\epsilon$ is the vacuum dielectric constant, and $\epsilon_0$ is the silicon relative dielectric constant. The calculated depletion voltage is in the range of 34 to 67 V, approximately.

\begin{figure}[!htb]
\includegraphics[width=\hsize]{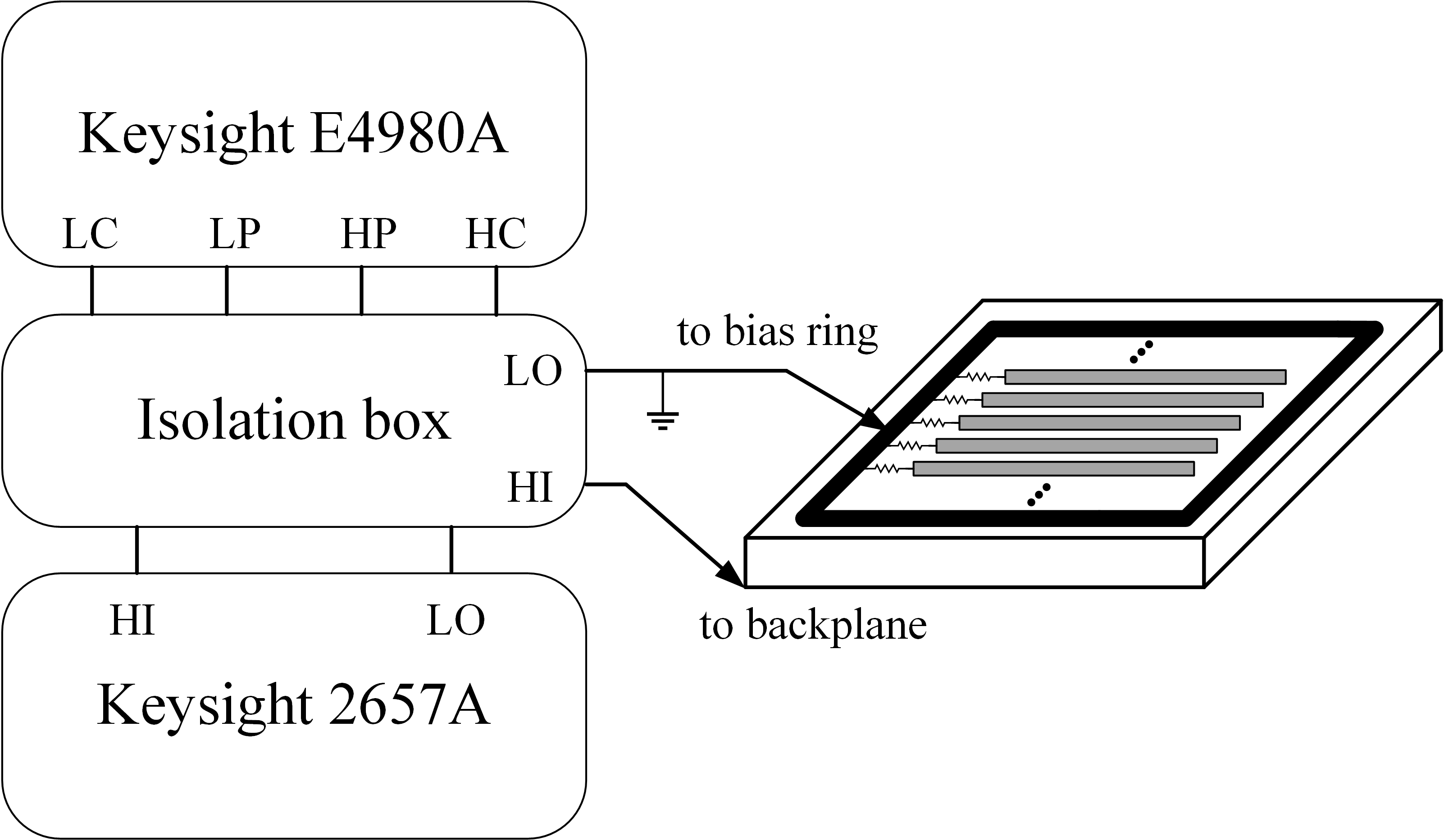}
\caption{Configuration for depletion voltage measurement}
\label{Fig_Configuration-for-the-depletion-voltage-measurement}
\end{figure}

To obtain the actual depletion voltage, we measured the bulk capacitance of the sensor by sweeping the bias voltage. Fig.~\ref{Fig_Configuration-for-the-depletion-voltage-measurement} shows the measurement setup. This test configuration uses a Keysight E4980A LCR meter to measure capacitances. The open and short corrections of the E4980A are implemented to compensate for any stray admittances and residual impedances, respectively. To make the sensor work properly, a bias voltage should be applied. Although the Keysight E4980A has a built-in DC bias of up to 40 V, we still need an independent source meter that can provide a DC voltage exceeding the estimated depletion voltage (i.e., ~67 V). A Keithley 2657A source meter is chosen. An isolation box (i.e., ~Keysight 16065A) is used to mix the DC bias (from the source meter) and the AC test signal (from the LCR) together, and then the outputs are connected to the sensor, which is located in the shielding chamber of a probe station.

As the reverse bias voltage increases, the bulk capacitance per unit area $C_{bulk}$ decreases to a constant when the maximum depletion depth is achieved. We rearrange Eq.~(3) in \cite{b16} as
\begin{equation}
\label{Eq_Depletion-voltage-B}
\frac{1}{C_{bulk}^2} = \left\{
        \begin{array}{cc}
            \frac{2V_{bias}}{q \epsilon_{si} |N_{effect}|} & V_{bias} \le V_{depl} \\
            &\qquad \qquad \qquad ,\\
            \frac{d_{depl}^2}{\epsilon_{si}^2} & V_{bias} > V_{depl}
        \end{array}
    \right.
\end{equation}
where $d_{depl}$ is the full depletion depth, $q$ is the electrical charge, $N_{effective}$ is the effective charge carrier density, and $\epsilon_{si}$ is the dielectric constant of the bulk silicon.

\begin{figure}[!htb]
\includegraphics[width=\hsize]{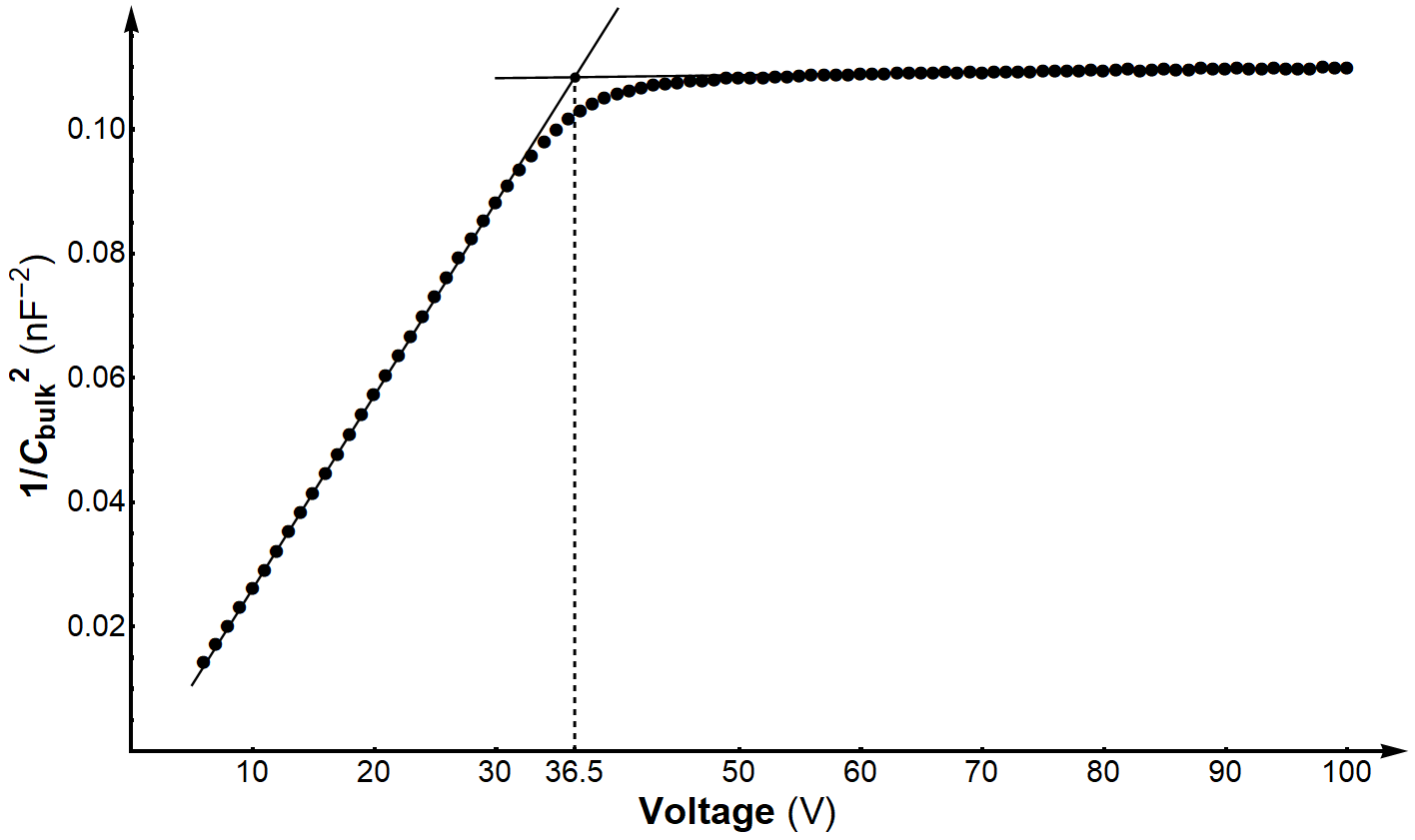}
\caption{Depletion voltage extraction}
\label{Fig_Depletion-voltage-extraction}
\end{figure}

A plot of $1/C_{bulk}^2$ verse $V_{bias}$ based on the measured data is shown in Fig.~\ref{Fig_Depletion-voltage-extraction}. We fit the slope and flat parts of the data points with a straight line. The depletion voltage can be extracted at the intersection point of these fitting lines, i.e.,~36.5 V. To reduce the charge collection time and the ballistic deficit~\cite{b17}, we often bias the sensor at a higher voltage (80 V), i.e., ~twice the depletion voltage.

\subsection{Bias resistance}

The bias resistor's functions are to bias each strip at a specific potential and to isolate different strips. The sensor uses a doped poly-silicon meandering line, i.e., ~poly-silicon resistor, to connect the strip's endpoint and the bias ring. It is incorrect to measure the bias resistance by just applying a voltage source and measuring the current at the terminals (i.e., ~the DC pad and the bias ring). In this incorrect circumstance, the main part of the testing current is bypassed through the DC pad (connected to the source and meter terminal), the corresponding strip, the backplane, other strips, and then to the bias ring (connected to the common ground terminal). Thus, the measured resistance is much lower than the actual value. This conclusion was confirmed by measurements.

\begin{figure}[!htb]
\includegraphics[width=\hsize]{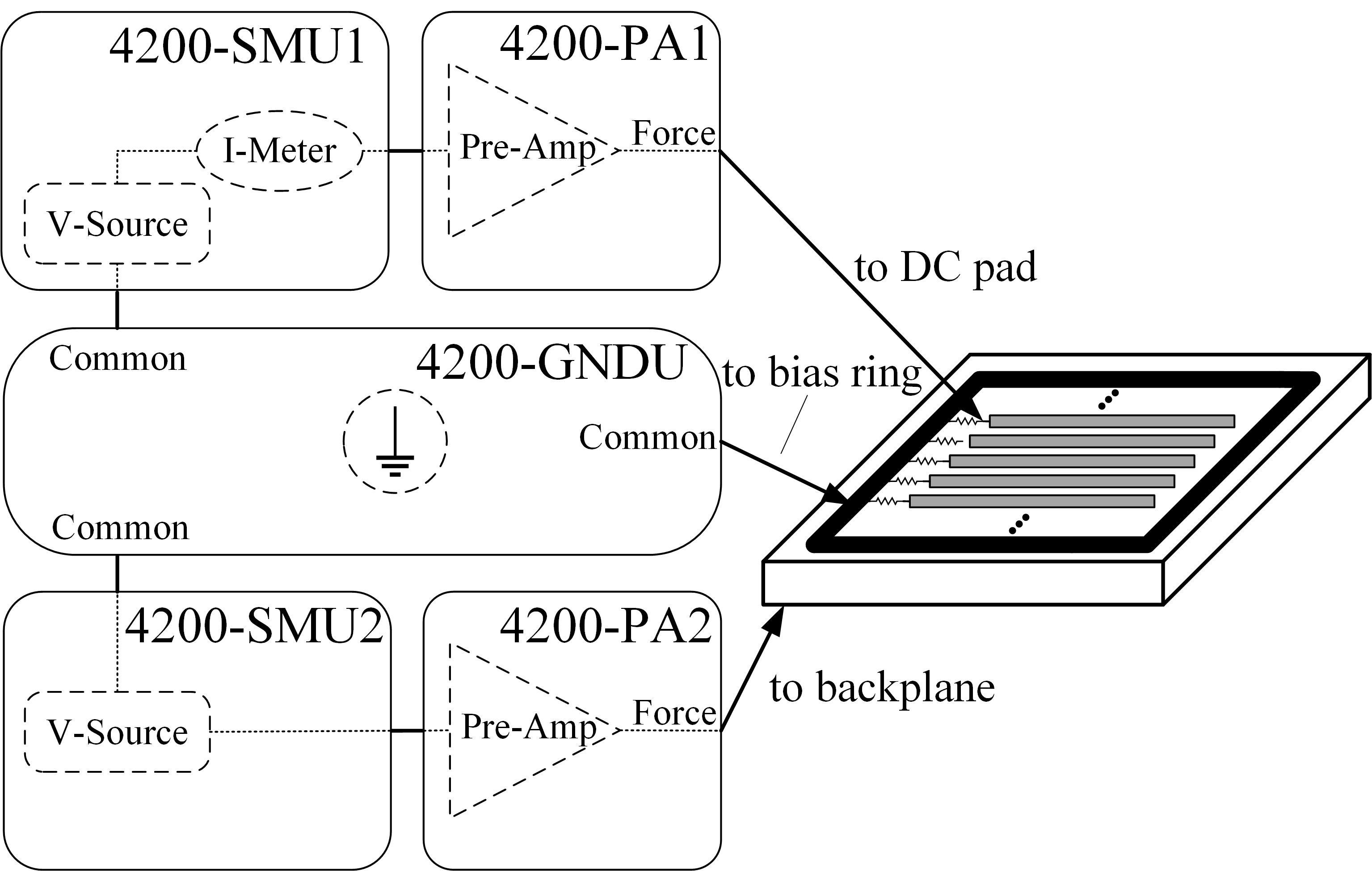}
\caption{Configuration for bias resistance measurement}
\label{Fig_Configuration-for-the-bias-resistance-measurement}
\end{figure}

The correct configuration to measure the bias resistance is shown in Fig.~\ref{Fig_Configuration-for-the-bias-resistance-measurement}. A multimodule integrated parameter analyzer, the Keithley 4200A-SCS, is adopted. We select three types of modules in the parameter analyzer: Source Measure Unit (SMU), Ground Unit (GNDU), and Preamplifier (PA). Next, we set up a local sense configuration~\cite{b18}. The SMU module is very flexible and can be configured by software as the source or the meter. In this configuration, the 2$^{\mathrm{nd}}$ SMU provides a bias voltage that depletes the substrate to block the bypass. The 1$^{\mathrm{st}}$ SMU applies a small testing voltage and measures its current at the same time, so the resistance can be calculated by the measured voltages and currents.

\begin{figure}[!htb]
\includegraphics[width=0.3\textwidth]{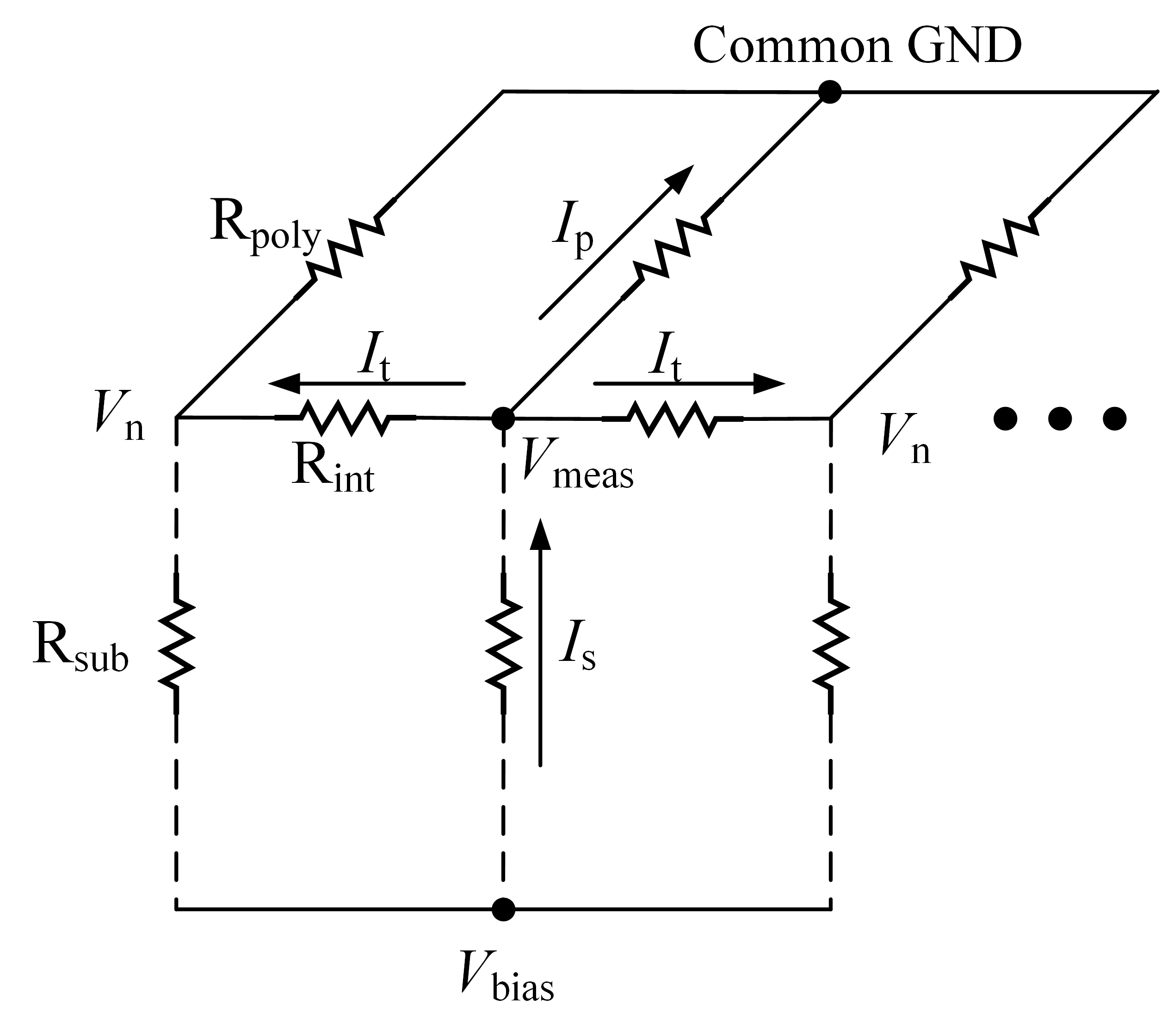}
\caption{SPICE model of bias resistance measurement}
\label{Fig_SPICE-model-of-the-bias-resistance-measurement}
\end{figure}

Owing to the on-chip parasitics, the silicon microstrip sensor can be modeled by a complicated SPICE RC network~\cite{b12}. The SPICE model at DC frequency, shown in Fig.~\ref{Fig_SPICE-model-of-the-bias-resistance-measurement}, is used to accurately analyze the calculated resistance. In this model, $R_{poly}$ represents the poly-silicon resistance, $R_{int}$ represents the interstrip resistance, and $R_{sub}$ represents the strip's backplane resistance. In addition, node potentials and path currents are marked in the figure. The unmarked current $I_{meas}$ is at node $V_{meas}$, and its direction points into the paper. We can easily get the relation as
\begin{equation}
\label{Eq_Bias_resistance-A}
I_{meas} + \frac{V_{bias}-V_{meas}}{R_{sub}} = \frac{V_{meas} }{R_{poly}} + 2 \times \frac{V_{meas}-V_n}{R_{int}}.
\end{equation}

The measured resistance is defined as the ratio of the voltage change and the current change:
\begin{equation}
\label{Eq_Bias_resistance-B}
R_{meas} = \bigtriangleup V_{meas}/\bigtriangleup I_{meas} = R_{poly} || R_{sub} || (0.5 R_{int}).
\end{equation}

\begin{figure}[!htb]
\includegraphics[width=\hsize]{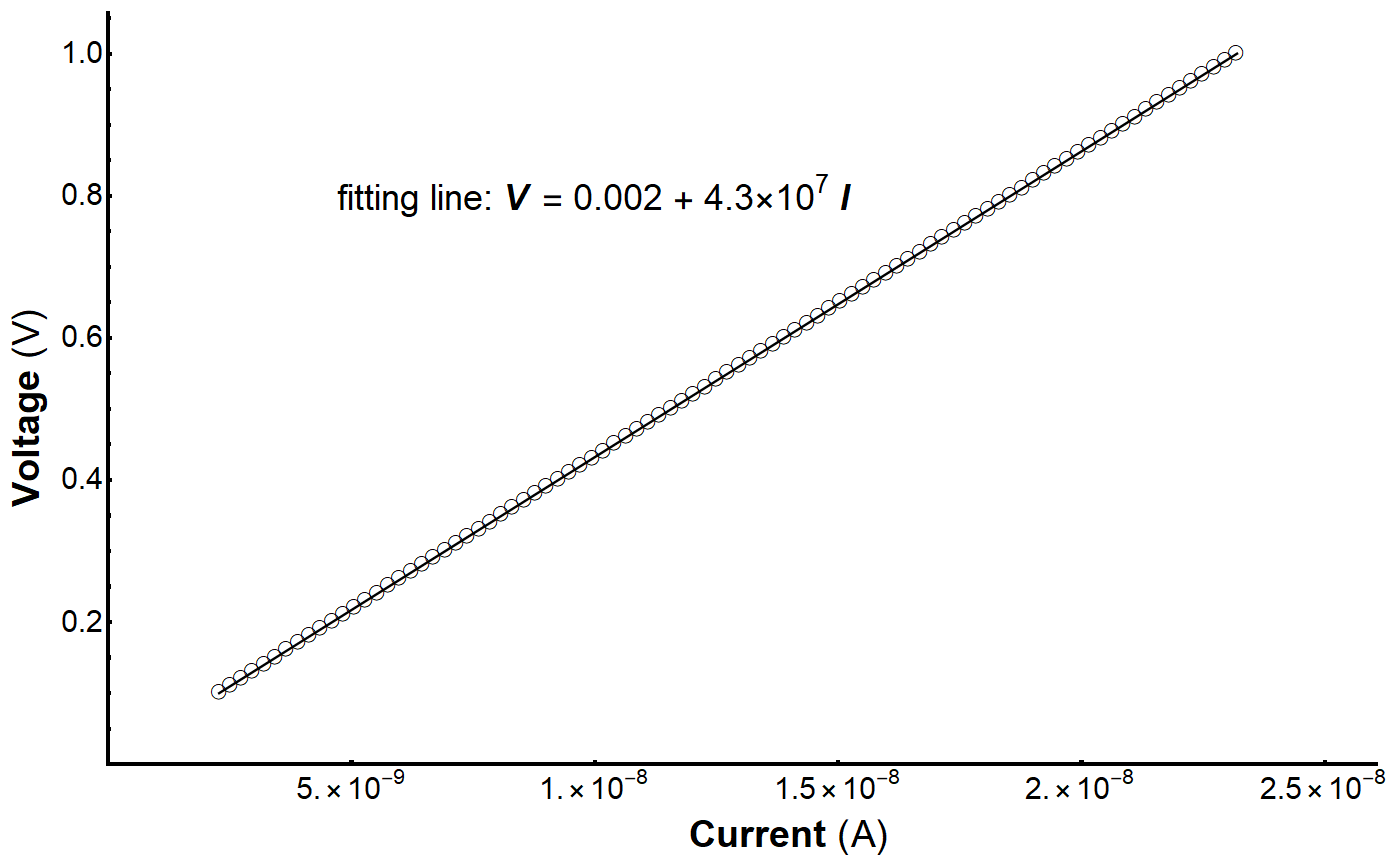}
\caption{Measurement result of bias resistance}
\label{Fig_Measurement-result-of-the-bias-resistance}
\end{figure}

Because $R_{sub}$ and $R_{int}$ are on the order of gigaohms while $R_{poly}$ is on the order of megaohms, the measured resistance is approximately equal to the poly-silicon resistance. As shown in Fig.~\ref{Fig_Measurement-result-of-the-bias-resistance}, the measured V--I curve can be well-fitted by a straight line, and the slope of the fitting line is 43 M$\Omega$, which represents the bias resistance. The measured bias resistance seems reasonable because it lies in the middle of the range of 20 to 60 M$\Omega$ from the manufacturer's specifications.

\subsection{Metal strip resistance}

The metal strip floats above the silicon oxide layer, couples the charge signal generated in the underlying depletion area, and transmits the signal to the external preamplifier in the readout ASICs.  Owing to its tiny width and thinness, the resistance of the long metal strip cannot be sufficiently low, and thus it contributes a thermal noise in series with the preamplifier's input. According to Fig.~4 in \cite{b19}, as the length of the ladder increases, the equivalent noise charge (ENC) owing to the metal strip resistance gradually becomes the dominant noise, so it is necessary to accurately measure the resistance.

\begin{figure}[!htb]
\includegraphics[width=\hsize]{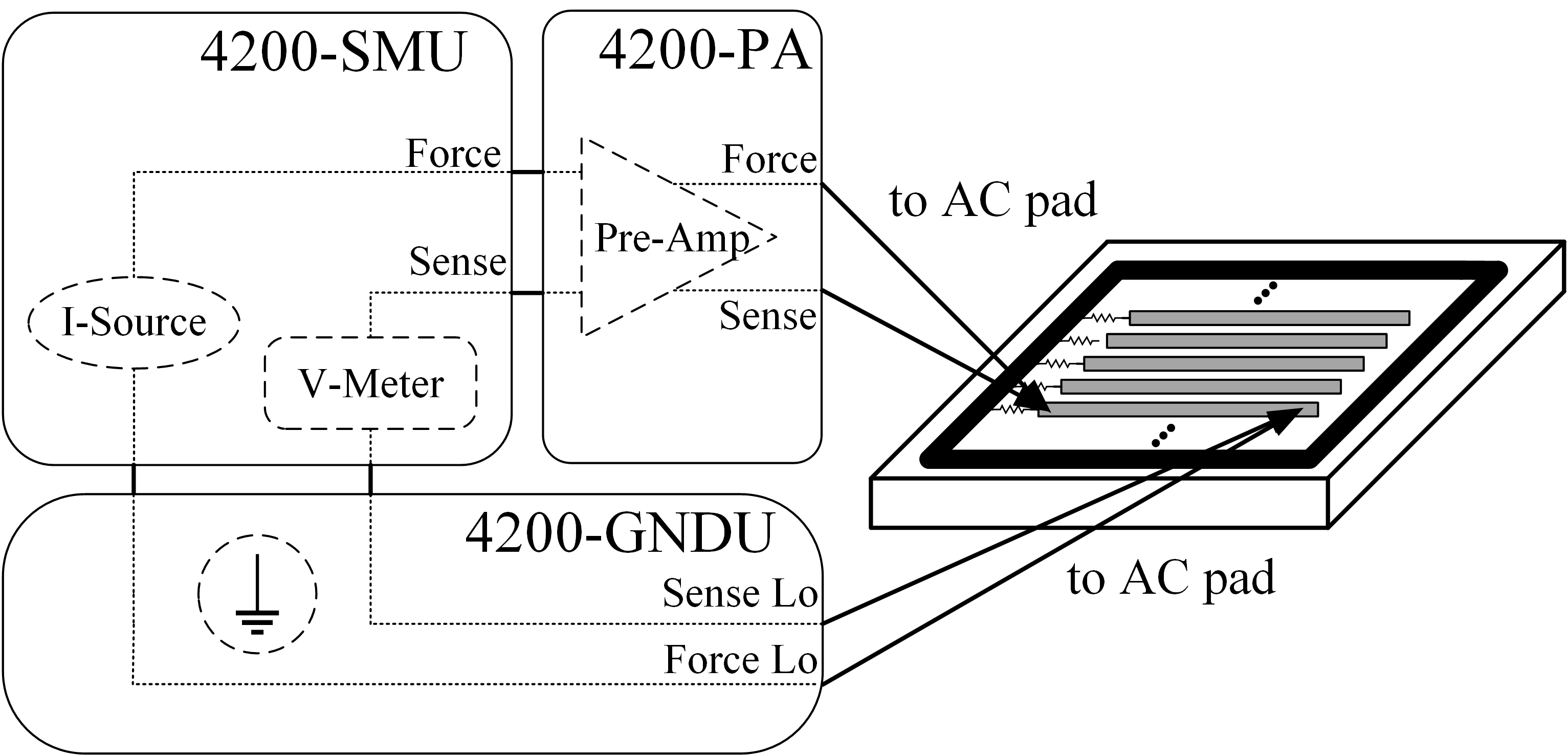}
\caption{Configuration for metal strip resistance measurement}
\label{Fig_Configuration-for-the-metal-strip-resistance-measurement}
\end{figure}

Because the metal strip resistance is relatively low (about tens of ohms), a full-Kelvin remote sensor measurement \cite{b18} should be adopted to eliminate the cable resistance and the probe-to-pad resistance. The measurement configuration is shown in Fig.~\ref{Fig_Configuration-for-the-metal-strip-resistance-measurement}. Using a set of testing cables, current is forced through the left AC pad, the metal strip, and then the right AC pad. By contrast, the voltage across the metal strip is measured using a set of sensing cables. The two sets of cables are only allowed to be connected to the corresponding AC pad by the probe tips. The testing current sweeps from 1 mA to 100 mA in steps of 1 mA, so 100 resistance samples are obtained by calculating the ratio of the measured voltage and the corresponding current. We randomly select three  metal strips (No.~100, No.~103, and No.~105), and the measured resistances are 30.292$\pm$0.018 $\Omega$, 30.300$\pm$0.018 $\Omega$, and 30.305$\pm$0.007 $\Omega$, respectively.

\subsection{Total leakage current}

According to the sensor's structure and working mechanism, the signal is inherently added to the leakage current, so the leakage fluctuation will inevitably affect the sensor's signal-to-noise ratio (S/N) and energy resolution. Thus, the leakage current is one of the main noise resources in the sensor and is often taken as a primary measure of the sensor's quality.

\begin{figure}[!htb]
\includegraphics[width=\hsize]{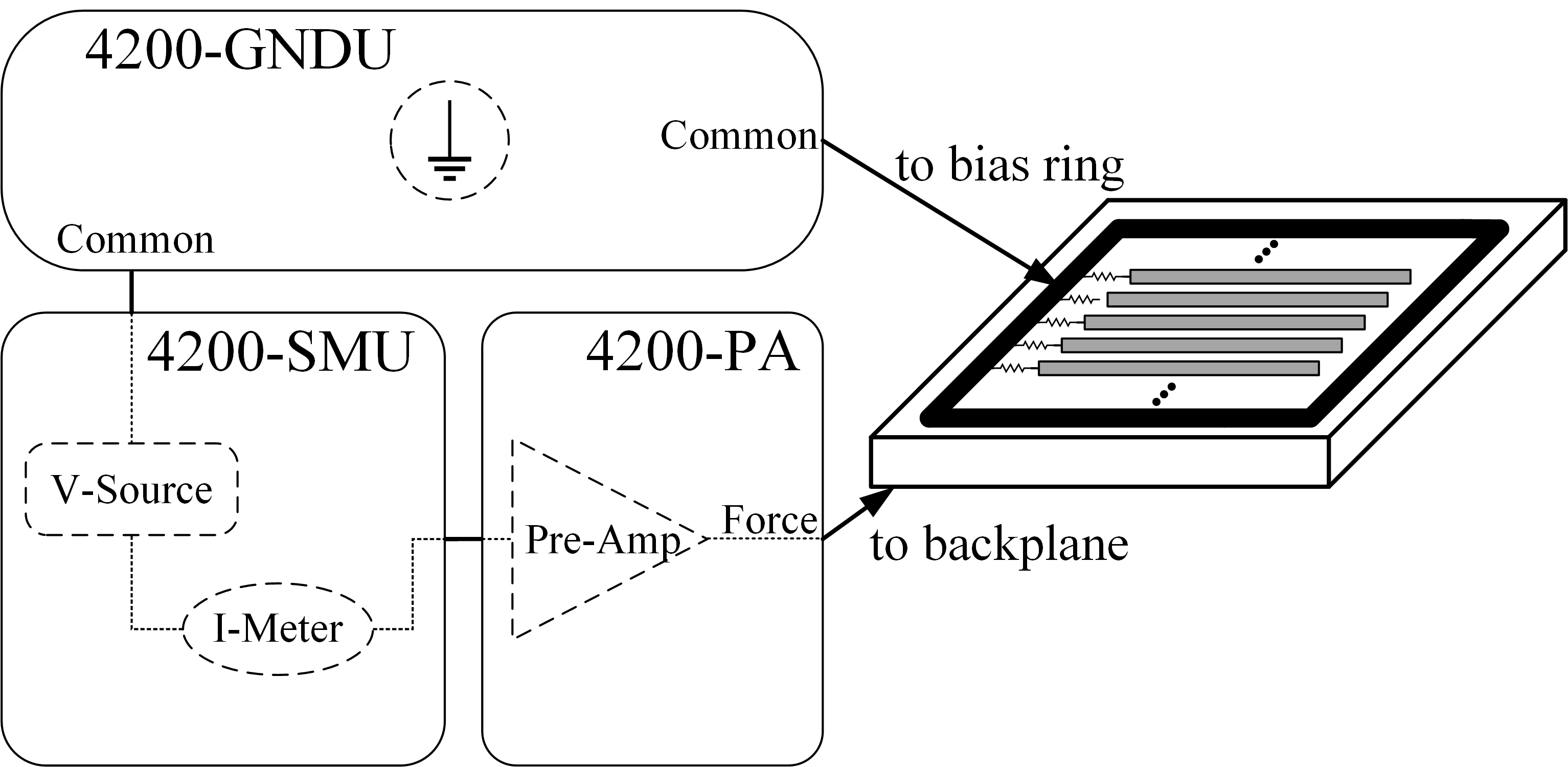}
\caption{Configuration for total leakage current measurement}
\label{Fig_Configuration-for-the-total-leakage-current-measurement}
\end{figure}

The sensor's total leakage current is equal to the sum of all strips and can represent the average leakage level. We can measure the total leakage current by the configuration shown in Fig.~\ref{Fig_Configuration-for-the-total-leakage-current-measurement}. In order to protect the sensor against damage, we activate a dual sweeping bias voltage from 1 V to 100 V and then from 100 V to 1 V, with a step of 1 V. In addition, triaxial cables are used to connect the parameter analyzer and the sensor to guarantee measurement accuracy. In addition, the leakage current is sensitive to environmental conditions, so we control the clean room's temperature and humidity at 20$\pm$1$^\circ$C and 40$\pm$5\%, respectively.

\begin{figure}[!htb]
\includegraphics[width=\hsize]{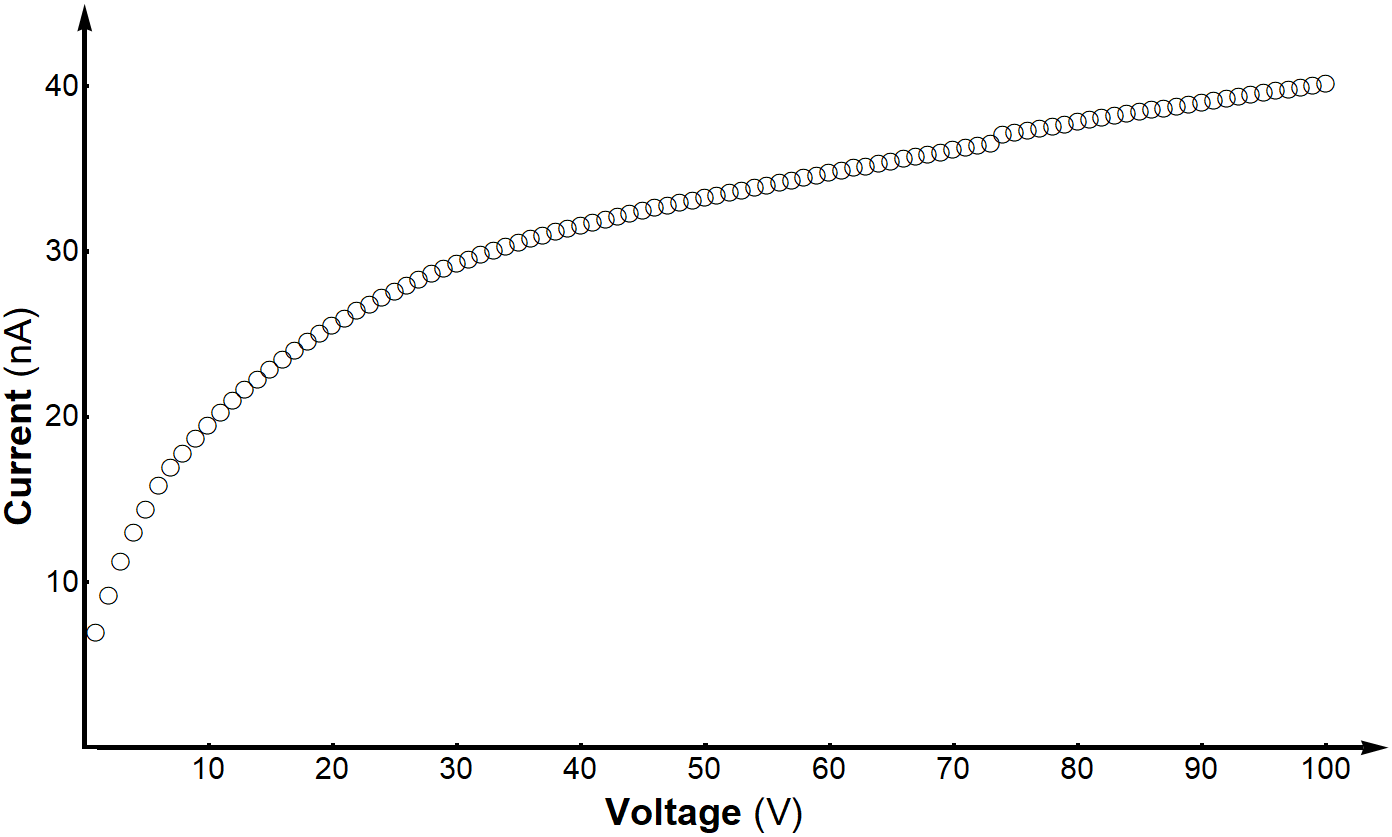}
\caption{Measurement result of total leakage current}
\label{Fig_Measurement-result-of-the-total-leakage-current}
\end{figure}

The measured I--V characteristics of the total leakage current are shown in Fig.~\ref{Fig_Measurement-result-of-the-total-leakage-current}. As the reverse bias voltage increases, the total leakage current increases accordingly and gradually reaches saturation. It can be observed that the sensor's total leakage current is very low. The average current density is on the order of 0.3 nA$/$cm$^2$ and 0.4 nA$/$cm$^2$ at the depletion voltage and operating voltage, respectively.

\subsection{Strip leakage current}

Owing to fabrication defects and manipulation damage, a few implant strips may appear to have a much higher leakage current than the average leakage level. The value of the strip leakage current is used to identify these bad strips from hundreds of good ones. To accurately measure the strip leakage current, there are two challenges. First, each implant strip of our sensor is physically connected to the bias ring through a poly-silicon resistor, so the leakage currents of all strips converge to the bias ring, which is connected to the common ground. Thus, the device-under-test (DUT) strip's leakage current should be separated from the total leakage current. Second, because the strip leakage current is on the order of tens to hundreds of picoamperes, it requires a much lower system noise in the measurement.

\begin{figure}[!htb]
\includegraphics[width=\hsize]{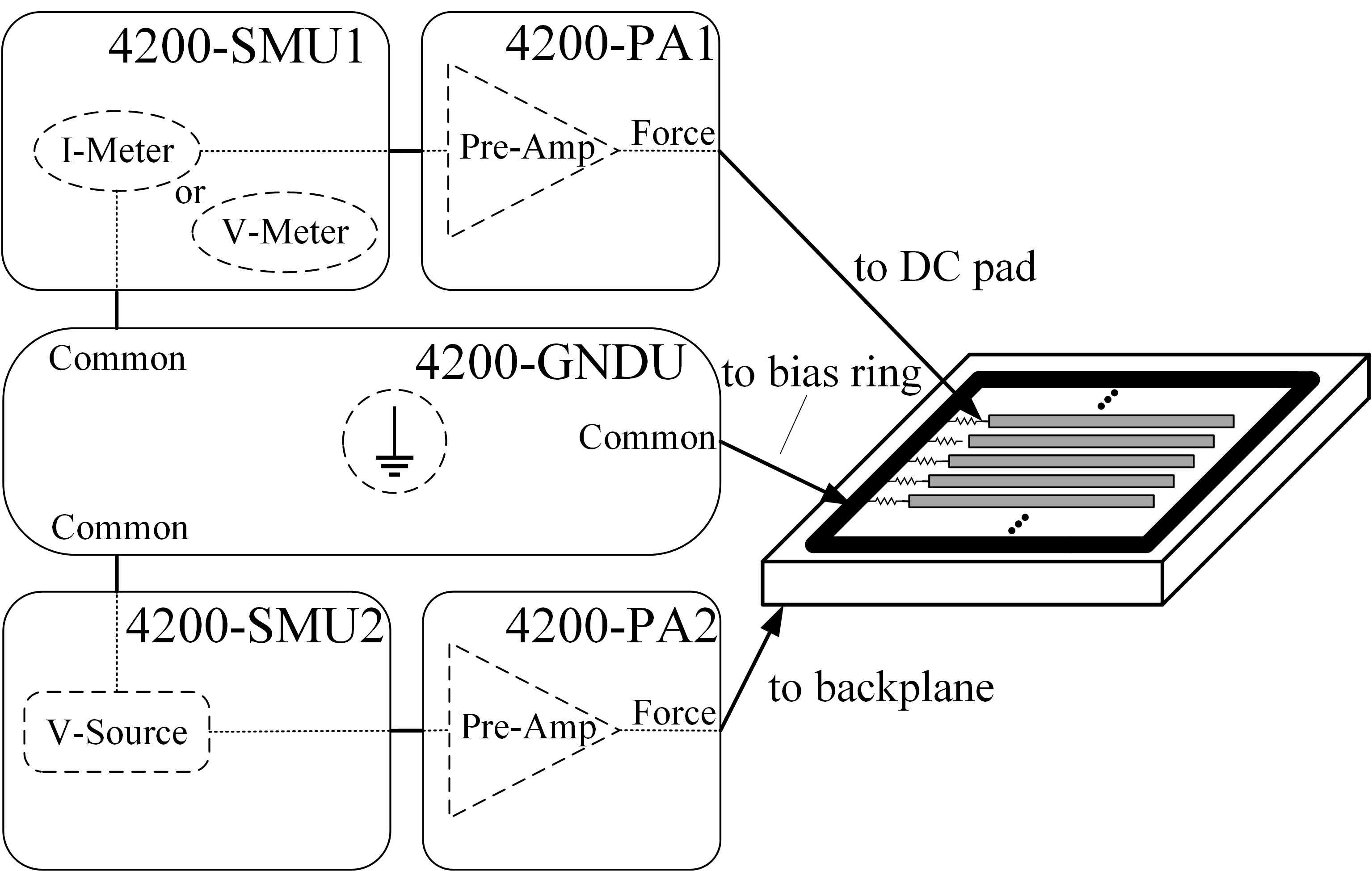}
\caption{Configuration for strip leakage current measurement}
\label{Fig_Configuration-for-the-strip-leakage-current-measurement}
\end{figure}

\begin{figure}[!htb]
\includegraphics[width=\hsize]{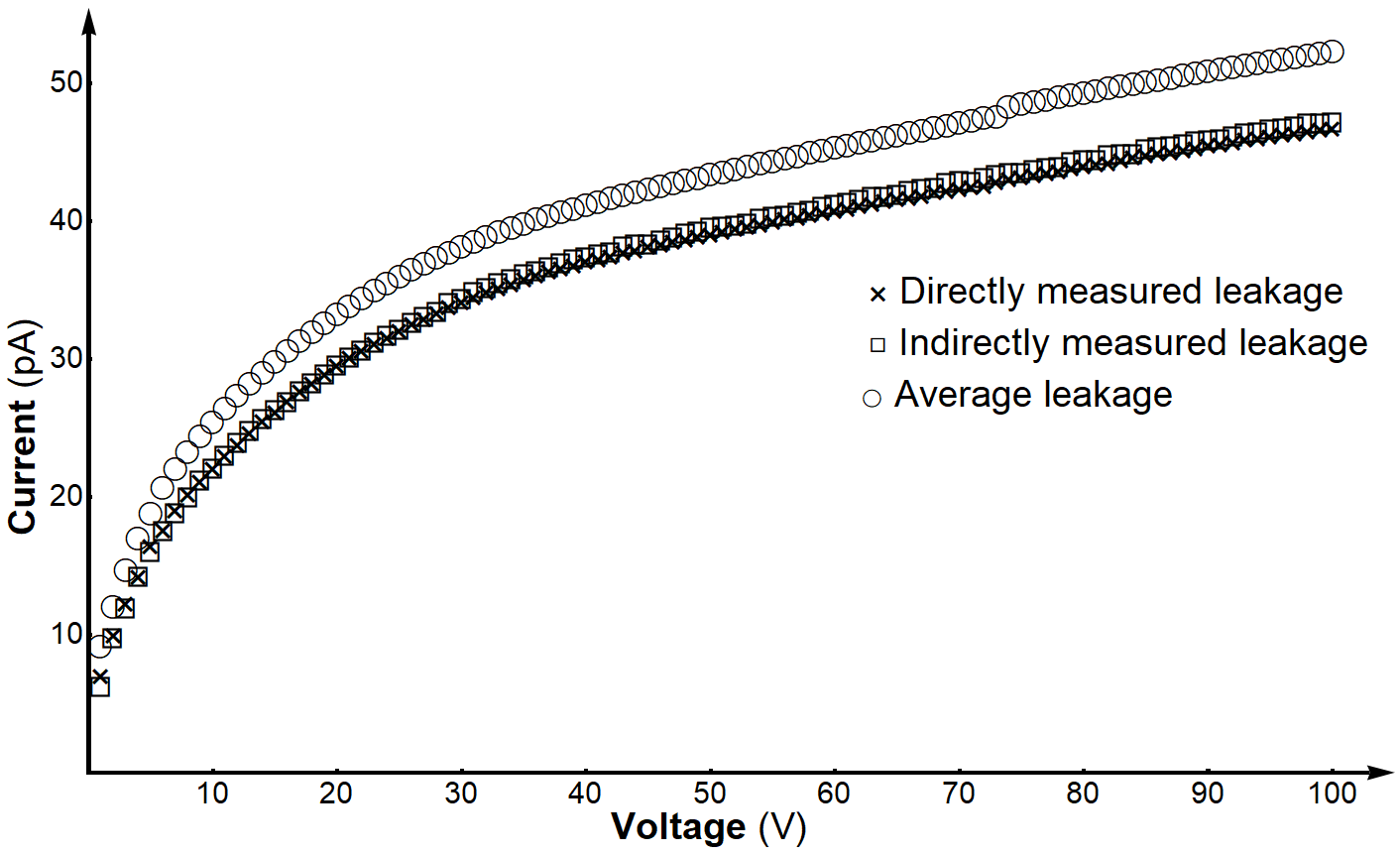}
\caption{Measurement result of strip leakage current}
\label{Fig_Measurement-result-of-the-strip-leakage-current}
\end{figure}

Two measurement configurations are proposed in Fig.~\ref{Fig_Configuration-for-the-strip-leakage-current-measurement}. To directly measure the strip leakage current, we configure the 1$^{\mathrm{st}}$ SMU as a current meter (i.e.,~select I-Meter in this figure). In this configuration, the DUT strip's leakage current flows into the current meter, while the remaining strips' leakage currents flow into the common ground. These two parts of the total leakage current are separated by the DUT strip's poly-silicon resistor that is about tens of megaohms. The directly measured I--V curve is shown in Fig.~\ref{Fig_Measurement-result-of-the-strip-leakage-current}, and it is similar to the total leakage current in Fig.~\ref{Fig_Measurement-result-of-the-total-leakage-current} but much smaller.

We can also measure the strip’s leakage current indirectly when the 1$^{\mathrm{st}}$ SMU is configured as a voltage meter (i.e.,~select V-Meter in Fig.~\ref{Fig_Configuration-for-the-strip-leakage-current-measurement}). In this configuration, the poly-silicon resistor and the strip's substrate resistor form a simple voltage divider, i.e.,
\begin{equation}
\label{Eq_Strip-leakage-Current-A}
V_{measure} = V_{sweep}\times\frac{R_{poly}}{R_{poly}+R_{sub}},
\end{equation}
where $V_{measure}$ is the voltage measured by the voltage meter, and $V_{sweep}$ is the sweep voltage applied by the voltage source.

In the previous subsection, there was a small voltage drop between the DUT strip and its neighboring strips, so Eq.~(\ref{Eq_Bias_resistance-B}) includes the interstrip resistance. The indirect configuration here is similar to the configuration for the bias resistance measurement but without a testing signal injected into the DC pad. Because there is no such voltage drop, the input resistance when looking into the DC pad with respect to the common ground is
\begin{equation}
\label{Eq_Strip-leakage-Current-B}
R_{dc} = R_{poly} || R_{sub}.
\end{equation}
Because the interstrip resistance is very large compared with $R_{poly} || R_{sub}$, it is reasonable to assume that $R_{dc}$ is equal to the measured bias resistance.

With Eq.~(\ref{Eq_Strip-leakage-Current-A}) and Eq.~(\ref{Eq_Strip-leakage-Current-B}), we can get
\begin{equation}
\label{Eq_Strip-leakage-Current-C}
R_{poly}=\frac{R_{dc}V_{sweep}}{V_{sweep}-V_{poly}}\;, \;R_{sub}=\frac{R_{dc}V_{sweep}}{V_{poly}}.
\end{equation}

With Eq.~(\ref{Eq_Strip-leakage-Current-C}), the strip leakage current is defined and calculated by
\begin{equation}
\label{Eq_Strip-leakage-Current-D}
I_{strip}=\frac{V_{sweep}}{R_{poly}+R_{sub}}=\frac{V_{poly}(1-V_{poly}/V_{sweep})}{R_{dc}}.
\end{equation}

The indirectly measured result is shown in Fig.~\ref{Fig_Measurement-result-of-the-strip-leakage-current}, and the average leakage current calculated from the total leakage current is also added. As shown in this figure, the directly and indirectly measured strip leakage currents match each other well, and both currents are close to the average leakage level.

\subsection{Coupling capacitance}

\begin{figure}[!htb]
\includegraphics[width=0.3\textwidth]{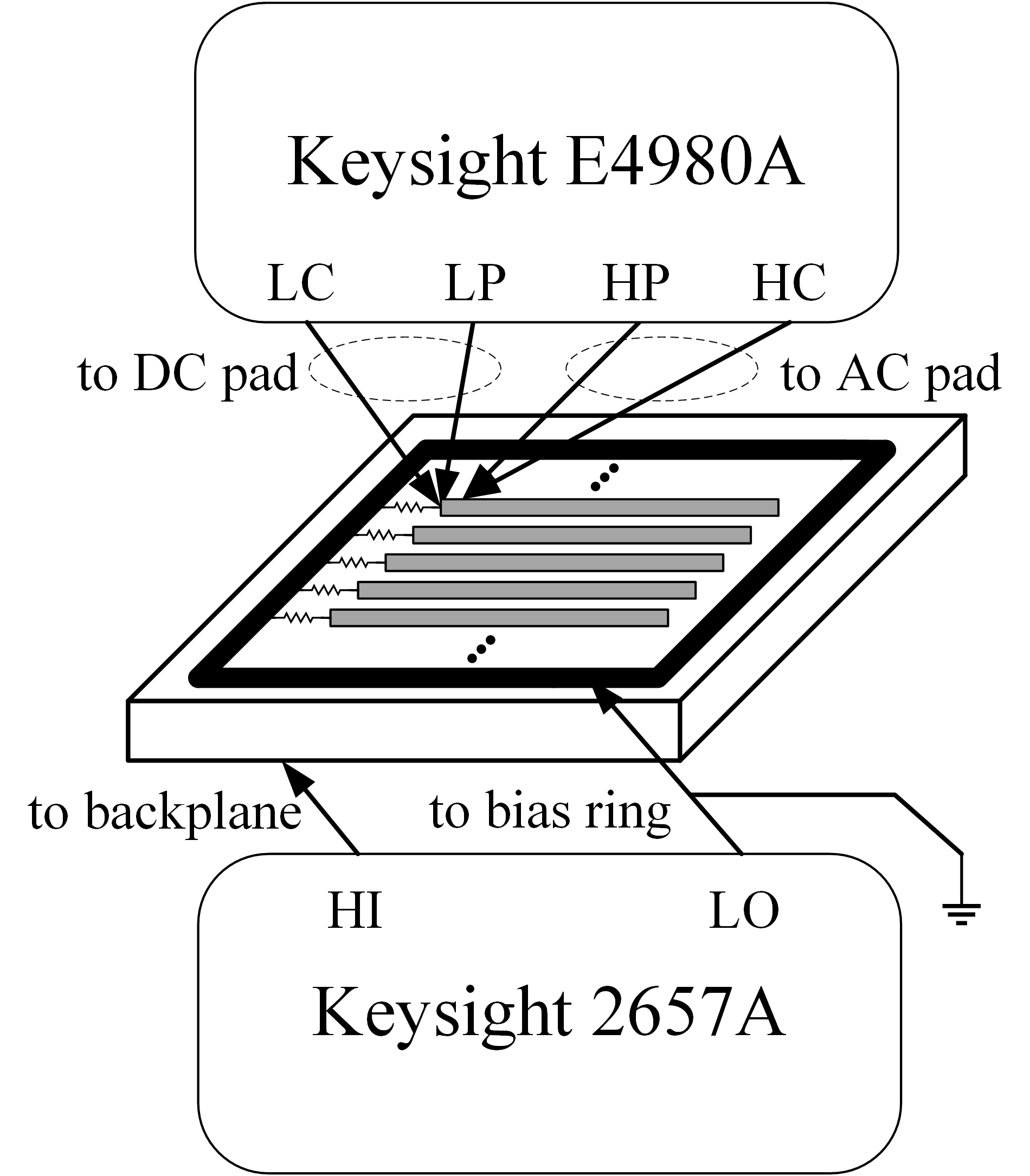}
\caption{Configuration for coupling capacitance measurement}
\label{Fig_Configuration-for-the-coupling-capacitance-measurement}
\end{figure}

The coupling capacitor in the silicon microstrip sensor has two main functions. First, it couples the charge in the implant strip to the metal strip that is connected to the readout ASICs. Second, the coupling capacitor can isolate the readout ASICs from the sensor's bias voltage and leakage current. In our sensor, the coupling capacitor is implemented as a ``metal--insulator--semiconductor" structure, and two terminals of the capacitor are the DC pad and AC pad of the DUT strip. Fig.~\ref{Fig_Configuration-for-the-coupling-capacitance-measurement} shows the measurement configuration. The high ports (HP and HC) and low ports (LP and LC) of the Keysight E4980A LCR meter are connected to the coupling capacitor's two terminals, while a Keithley 2657A provides a working bias voltage. The open and short corrections of the E4980A are implemented to guarantee the measurement accuracy.

\begin{figure}[!htb]
\includegraphics[width=\hsize]{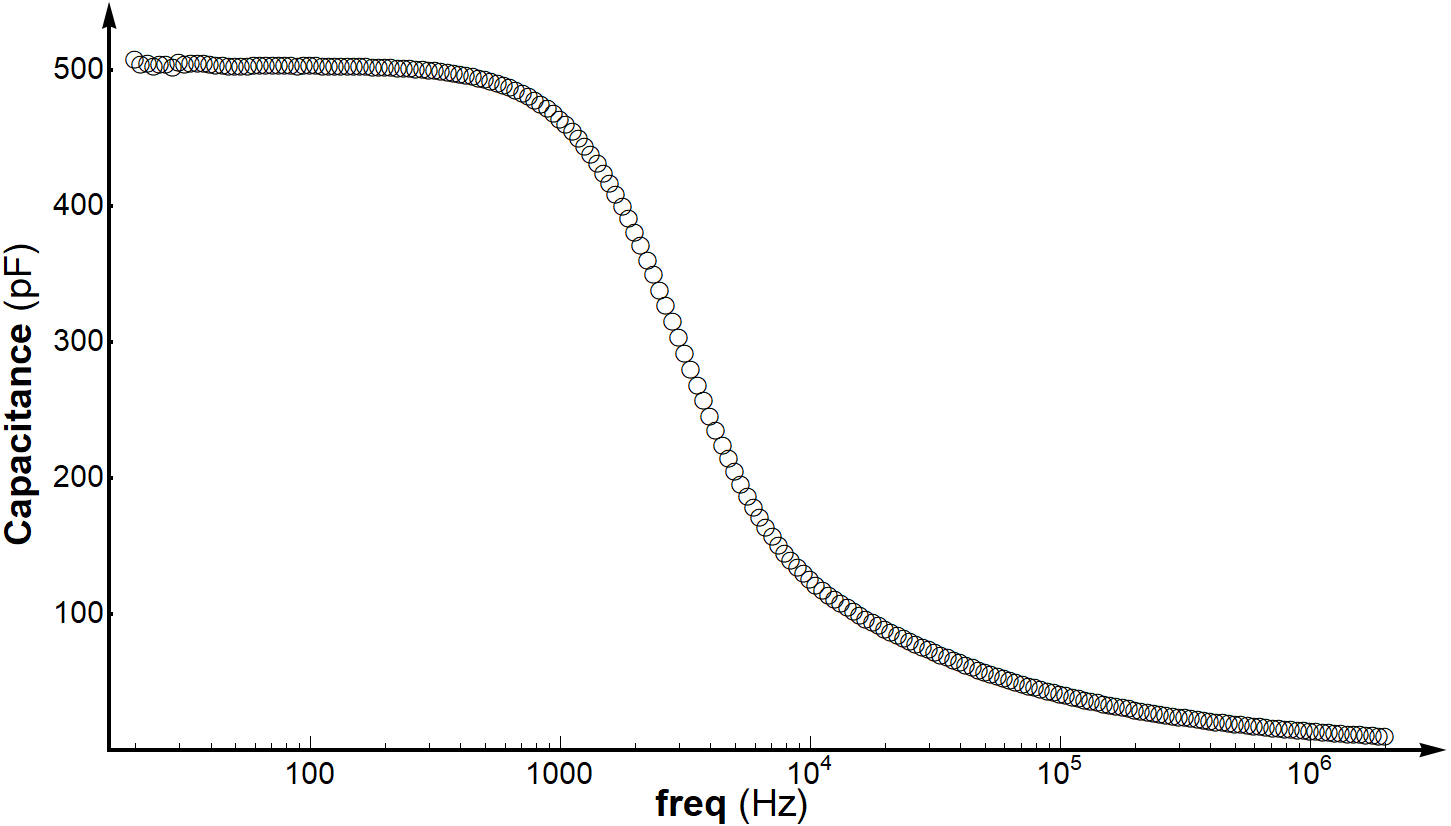}
\caption{Measurement result of coupling capacitance}
\label{Fig_Measurement-result-of-the-coupling-capacitance}
\end{figure}

The measured C--f curve is shown in Fig.~\ref{Fig_Measurement-result-of-the-coupling-capacitance}. In the figure, the curve indicates a strong dependence on the frequency. When the frequency is lower than 500 Hz, the measured capacitance is high and relatively stable. As the frequency increases, the capacitance drops dramatically. When the frequency is higher than 2 MHz, which is beyond our testing scope, the capacitance seems to approach zero.

\begin{figure}[!htb]
\includegraphics[width=\hsize]{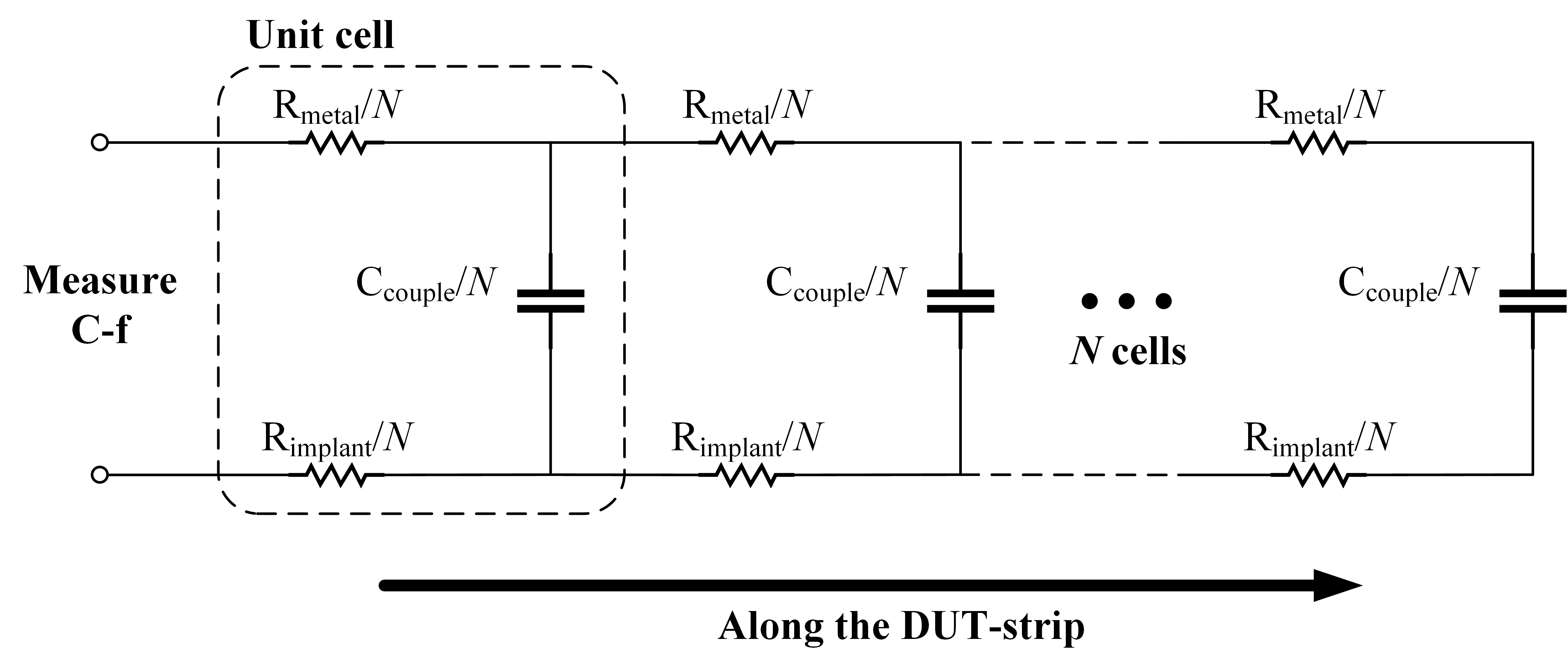}
\caption{SPICE model of coupling capacitance measurement}
\label{Fig_SPICE-model-of-the-coupling-capacitance-measurement}
\end{figure}

To quantitatively analyze the measured C--f curve, we divide the DUT strip's SPICE model into $N$ unit cells, as shown in Fig.~\ref{Fig_SPICE-model-of-the-coupling-capacitance-measurement}. Each unit cell comprises three components: $R_{metal}/N$ and $R_{implant}/N$ represent the unit cell's metal strip resistance and implant strip resistance, respectively, and $C_{couple}/N$ is the unit cell's coupling capacitance. When the unit cell is treated as a two-port network, we can easily obtain its transmission matrix as
\begin{equation}
\label{Eq_Coupling-capacitance-A}
T_{cell}=
  \left(
    \begin{array}{cc}
      1+\frac{j\omega C_{couple}(R_{metal}+R_{implant})}{N^2} & \frac{R_{metal}+R_{implant}}{N} \\
      & \\
      \frac{j\omega C_{couple}}{N} & 1
    \end{array}
  \right).
\end{equation}
Because the DUT strip's SPICE model is a cascade connection of $N$ unit cells, the transmission matrix of the DUT strip equals the product of all cells' transmission matrices:
\begin{equation}
\label{Eq_Coupling-capacitance-B}
T_{strip}=T_{cell}^N=
  \left(
    \begin{array}{cc}
      T_1 & T_2\\
      T_3 & T_4
    \end{array}
  \right).
\end{equation}
The capacitor looking into the DUT strip can be calculated as
\begin{equation}
\label{Eq_Coupling-capacitance-C}
C_{meas}=\frac{T_3}{j\omega T_1}
\end{equation}
Although the analytical expression is complicated, we can give the limit value when the frequency is zero and infinity as
\begin{equation}
\label{Eq_Coupling-capacitance-D}
\lim_{\omega \to 0}C_{meas}=C_{couple}\;, \;\lim_{\omega \to \infty}C_{meas}=0.
\end{equation}
The limit values well match the shape of the C--f curve in Fig.~\ref{Fig_Measurement-result-of-the-coupling-capacitance} and demonstrate that the coupling capacitance of the DUT strip can be extracted from the measured curve at low frequencies. In our experiment, the extracted coupling capacitance of the DUT strip is 508 pF.

\subsection{Interstrip capacitance}
The interstrip capacitance is the main part of the total capacitance seen by the preamplifier input. According the product information of a typical readout ASIC~\cite{b20}, ENC induced by the ASIC is in proportion to the input capacitance, so the interstrip capacitance should be reduced to suppress noise. By contrast, the interstrip capacitance should be increased to avoid a charge loss in the process of charge division. By measuring the interstrip capacitance, we can evaluate the noise induced by ASIC and the charge collection efficiency.

\begin{figure}[!htb]
\includegraphics[width=0.3\textwidth]{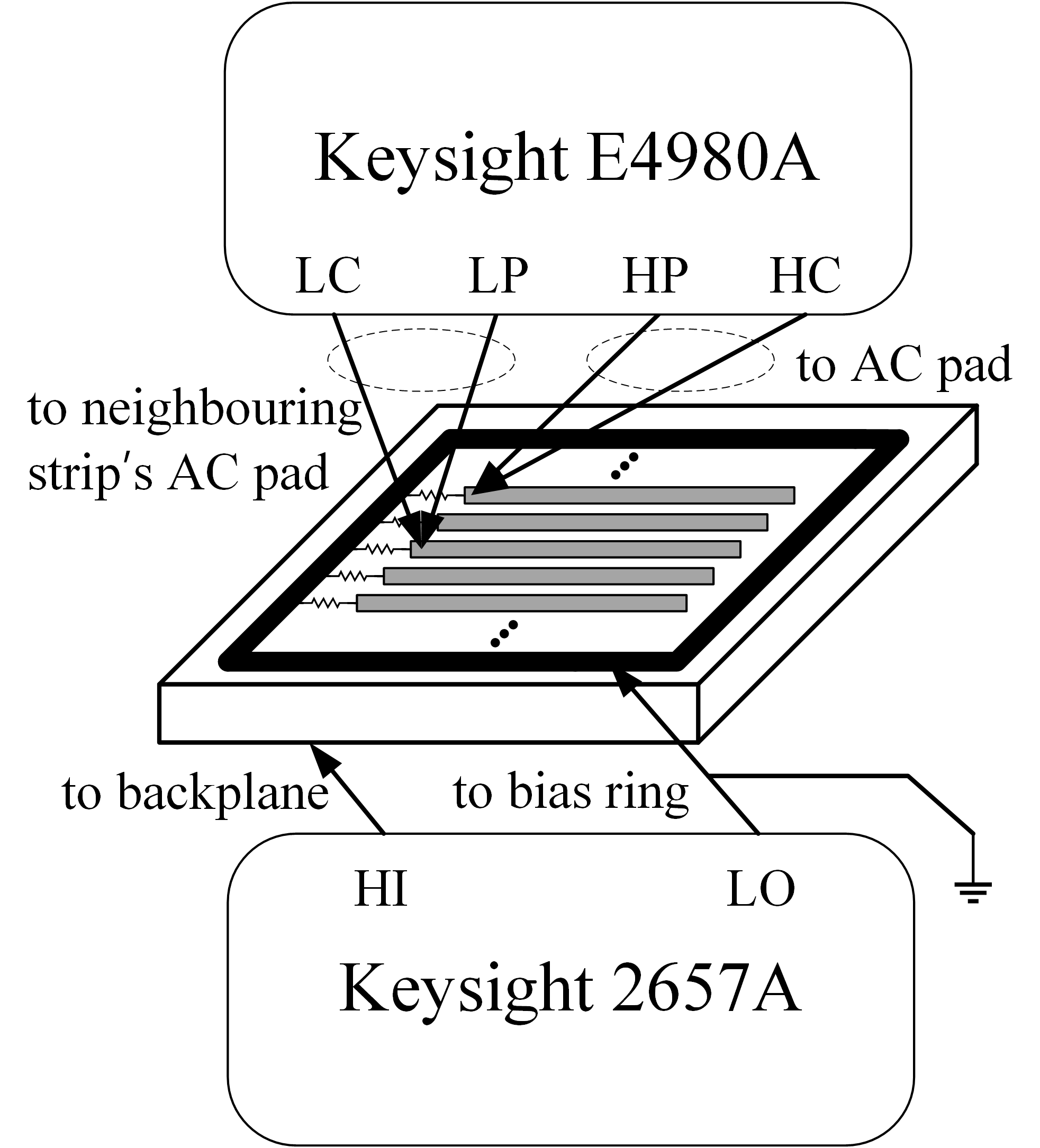}
\caption{Configuration for interstrip capacitance measurement}
\label{Fig_Configuration-for-the-interstrip-capacitance-measurement}
\end{figure}

The measurement configuration is shown in Fig.~\ref{Fig_Configuration-for-the-interstrip-capacitance-measurement}. The high ports (HP and HC) of the Keysight E4980A LCR meter are connecting to the DUT strip's AC pad, and the low ports (LP and LC) are connected to its neighboring strip's AC pad. The open and short corrections are used. To guarantee the sensor is in working condition, a bias voltage is applied by a Keithley 2657A.

\begin{figure}[!htb]
\includegraphics[width=\hsize]{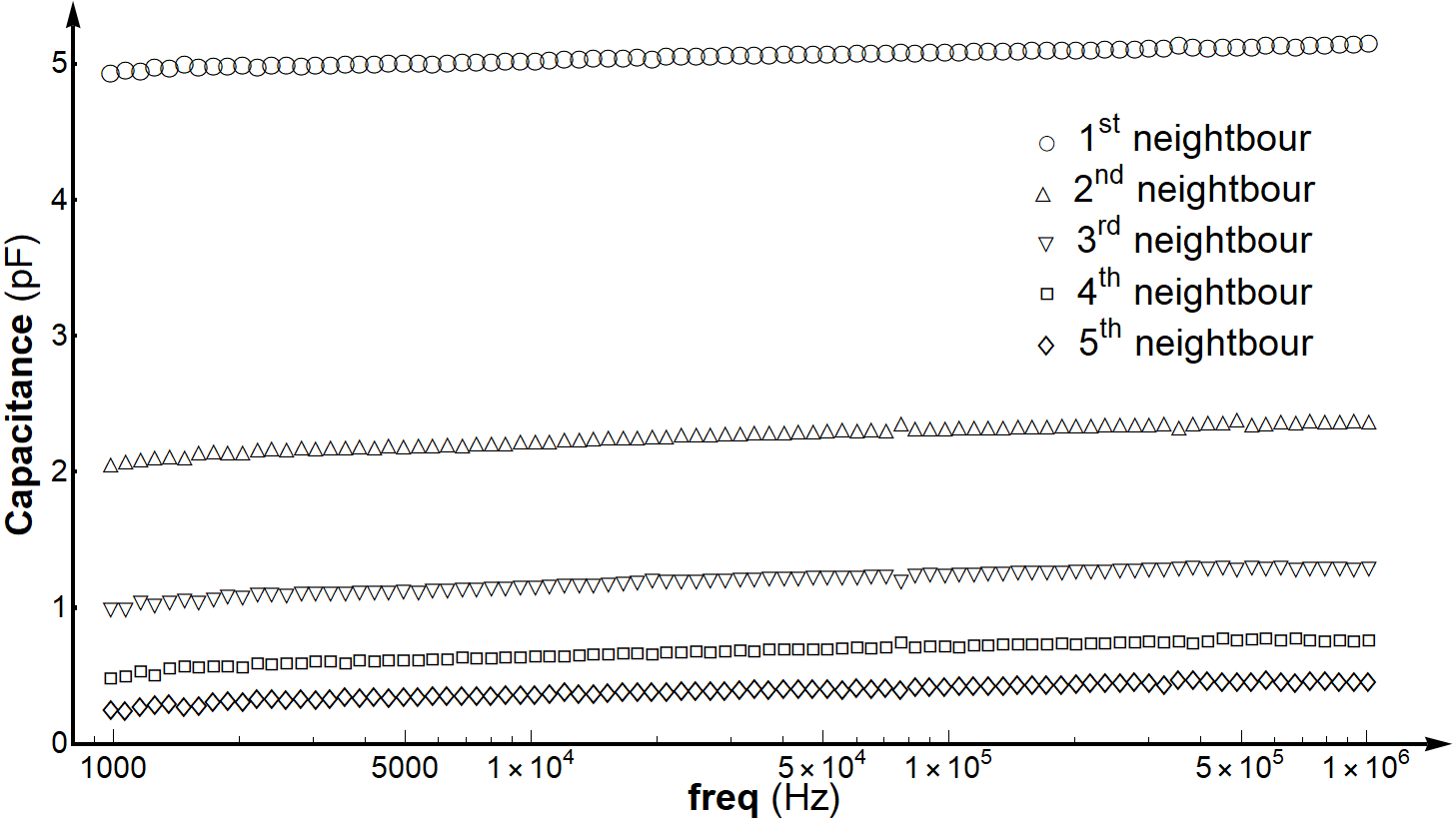}
\caption{Measurement result of interstrip capacitance}
\label{Fig_Measurement-result-of-the-interstrip-capacitance}
\end{figure}

Because the interstrip capacitance is very low (only a few picofarads), the impedance at low frequencies is very high. Thus, it is difficult to accurately measure the interstrip capacitance at low frequencies. Actually, the measured C--f curve fluctuates dramatically at low frequencies. To show the measured data clearly, we draw the C--f curve from 1 kHz in Fig.~\ref{Fig_Measurement-result-of-the-interstrip-capacitance}. As shown in the figure, we find that the interstrip capacitance is independent of the frequency and drops as the neighboring strip is farther away from the DUT strip. In our experiment, the extracted interstrip capacitances between the DUT strip and its neighbors are 5.15 pF, 2.38 pF, 1.29 pF, 0.75 pF, and 0.44 pF.

\section{Conclusion}
A silicon microstrip sensor used in space astronomy was extensively characterized in this work. To determine the depletion voltage, we measured the bulk capacitance and then fitted the measured data to extract the depletion voltage. In measuring the bias resistance, we first pointed out the necessity to apply a bias voltage in the test configuration. Second, a SPICE model was proposed to analyze the complex parasitics and deduce the expression of the measured result. Third, we linearly fitted the measured data and extracted the bias resistance according to the deduced expression. To accurately measure the metal strip resistance, we set up a full-Kelvin remote sensing configuration because the resistance was relatively low. The total leakage current was measured to evaluate the sensor's average leakage level. Direct and indirect measurement methods were adopted to characterize the strip leakage current, and the measured results matched each other well. In measuring the coupling capacitance, the measured curve showed a strong dependence on the frequency. We analyzed the curve quantitatively based on the SPICE model of the strip and extracted the correct value for the coupling capacitance. Finally, a series of interstrip capacitances was measured between the DUT strip and its neighbor strips.

\end{document}